\documentclass[preprint,showpacs,pra,aps,superscriptaddress,amssymb]{revtex4-1}

\usepackage{amsmath}
\usepackage{amsfonts}
\usepackage{mathrsfs}
\usepackage{amssymb}
\usepackage{bm}
\usepackage{graphicx}
\usepackage{color}
\usepackage[dvipsnames]{xcolor}
\usepackage[colorlinks=true,linkcolor=blue,citecolor=blue,urlcolor=blue]{hyperref}

\begin{document}

\title{Exact analytical non-Hermitian formulation of the time evolution of decay of one and two identical quantum particles}

\author{Gast\'on Garc\'{\i}a-Calder\'on}
\email{gaston@fisica.unam.mx}
\affiliation{Instituto de F\'{\i}sica,
Universidad Nacional Aut\'onoma de M\'exico, 04510 Ciudad de M\'exico, M\'exico}

\begin{abstract}
An analytical solution to the time evolution of decay of one and two identical noninteracting
particles is presented using the formalism of resonant states. It is shown that the time-dependent
wave function and hence the survival and nonescape probabilities for the initial
state of a single particle and entangled symmetric and antisymmetric initial states of two
identical particles evolve in a distinctive form along the exponential and long-time nonexponential decaying regimes. In particular, for the last regime, they exhibit different inverse power of time behaviors.
\end{abstract}

\maketitle

%
\section{Introduction}

Historically, the notion of quantum decay was developed to explain  $\alpha$-decay in radioactive nuclei. In 1928, Gamow derived the analytical expression for the exponential decay law $\exp(-\Gamma t/\hbar)$, with $\Gamma$ the decay rate, by imposing purely outgoing boundary conditions to the solutions to the Schr\"odinger equation that describes the decaying system \cite{gamow28}. Typically this is modeled by a potential having a barrier where the particle is initially confined prior to decay by tunneling into open space. The approach by Gamow constituted one of the first theoretical treatments of an open quantum system. The outgoing  boundary conditions, however, imply the vanishing of the coefficients of the incoming waves that appear in the general solution to the Schr\"odinger equation outside the interaction region and lead, due to time-reversal considerations, to  discrete complex energy eigenvalues. One sees, therefore, that purely outgoing boundary conditions imply a non-Hermitian formulation of decay. Nevertheless, this approach has been very successful in describing particle decay as a tunneling process not only in nuclei \cite{gamow49}, but also in other realms of physics as, for example, electronic decay in double-barrier semiconductor heterostructures \cite{sakaki87} or atomic decay in ultracold traps \cite{serwane11}. In the energy domain, the vanishing of the incoming wave coefficients mentioned above corresponds to the complex poles of the $S$-matrix to the problem. This has led to formulations of nuclear reactions involving resonance expansions of the cross section \cite{siegert39,rosenfeld61}. It is worth noticing, however, that this provides a link between the energy and the time domains, corresponding respectively, to scattering and decay, and to a definition of resonances as an intrinsic property of open quantum systems.

Near the end of the fifties of last century  Khalfin demonstrated that if the energy spectra $E$ of the system is bounded by below, \textit{i.e.}, $E \in (0,\infty)$, the exponential decay law cannot hold at long times, behaving instead as an inverse power of time \cite{khalfin58}. Studies on the short-time behavior of decay predicted also a departure from the exponential decay law. This is related, however, to the existence of the energy moments of the Hamiltonian $H$ \cite{khalfin68,muga96}. Here short and long times refer to the time scale set by the lifetime of the decaying system. The short-time behavior has been the subject of much discussion, particularly in connection with the quantum Zeno effect \cite{sudarshan77,koshino05}. The experimental search of  departures from the exponential decay law remained elusive for decades \cite{norman,opal}. A few years ago, however, it was verified in the short-time regime using ultracold atoms \cite{raizen97} and more recently, in the long-time regime using organic molecules in solution \cite{monk06}. The quantum Zeno effect has also been observed in ultracold decaying systems \cite{raizen01}. These experimental results seem to contradict theoretical claims made in the seventies of last century, that argued that due to the influence of the measurement apparatus on the decay process, the exponential decay law should hold at all times \cite{siegert71,fonda78}.

One sees, therefore, that the theoretical description of the dynamics of particle decay by tunneling has become more complex than was expected in older times. In general, it seems to consist of three regimes: Exponential and non-exponential  at short and long times. Clearly the approach initiated by Gamow requires substantial modification to deal with the non-exponential contributions.  At the end of the 1960s and along the 1970s there were some relevant developments on the properties of resonance states in the energy domain that involved consideration of the analytical properties of the outgoing Green's function to the full problem \cite{newton}. This function provided a framework to study the issues of normalization and eigenfunction expansions involving resonant states. Here it is adopted the normalization procedure that arises from the residue at a complex pole of the outgoing Green's function \cite{gcp76} and resonance expansions using the Cauchy integral theorem \cite{gc76,gcb79}. The above considerations form the basis for the time-dependent treatment that I shall review below to describe the time evolution of decay of one and two particles.

It might be worthwhile to point out that the formulation considered here, involving the full hamiltonian $H$ to the problem, should be contrasted with approaches where the Hamiltonian is separated into a part $H_0$ corresponding to a closed system and a part $H_1$ responsible for the decay which is usually treated to some order of perturbation theory, as in the work by Weisskopf and Wigner to describe the decay (also exponential) of an excited atom interacting with a quantized radiation field, published as well in the old days of quantum mechanics \cite{wigner30}. This approximate approach has become a standard procedure for treating the class of decay problems where perturbation theory can be justified.

The paper is organized as follows.  Section 2 discusses the time evolution of decay of a single particle. Section 3 refers to the decay of symmetric and anti-symmetric identical particles and, finally, section 4 presents some concluding remarks.
\section{Decay of a single particle}

Consider a single particle confined at $t=0$ along the internal region of a real spherically symmetrical finite-range potential, \textit{i.e.}, $V(r)=0$ for $r>a$. For simplicity we consider $s$ waves and choose as units
$\hbar=2m=1$. The solution to the time-dependent Schr\"odinger equation in the radial variable $r$, as an initial
value problem, may be written at time $t > 0$ in terms of the retarded Green's function $g(r,r';t)$ of the problem as
\begin{equation}
\Psi(r,t)=\int_0^a {\! g(r,r^\prime,t)\Psi(r',0)\,\mathrm{d}r^\prime},
\label{1s}
\end{equation}
where $\Psi(r,0)$ stands for the arbitrary state initially confined within the internal interaction region.
Since the decay  refers to tunneling into the continuum, for the sake of simplicity it is assumed that the  potential does not possess bound nor antibound states. It is convenient to express the
retarded time-dependent Green's function in terms of the outgoing Green's function $G^+(r,r';k)$ of the problem.
Both quantities are related by a Laplace transformation. The Bromwich contour in the $k$ complex plane  corresponds to a hyperbolic contour along the first quadrant that may be deformed to a contour that goes from $-\infty$ to $\infty$ along the real $k$ axis,
\begin{equation}
g(r,r';t)={i \over 2 \pi} \int_{-\infty}^{\infty} G^+(r,r\,';k) {\rm e}^{-i k^2t} \,2kdk.
\label{74}
\end{equation}
This allows to make use of the resonant expansion of the outgoing Green's function \cite{gc10}
\begin{equation}
G^+(r,r\,';k) = \frac{1}{2k}\sum_{n=-\infty}^{\infty} \frac {u_n(r)u_n(r\,')}{k-\kappa_n}, \quad  (r,r')^{\dagger} \leq a
\label{9x}
\end{equation}
where the notation $(r,r')^\dagger$  means that the point $r=r^\prime=a$ is excluded in the above expansion (otherwise it diverges) and the set of functions $\{u_n(r)\}$ correspond to the resonant  states (also known as quasinormal modes) of the problem. They follow from the residues at the complex poles $\{\kappa_n\}$ which also provide its normalization condition \cite{gcp76,gc10}
\begin{equation}
\int_0^a u_n^2(r) dr + i\frac{u_n^2(a)}{2\kappa_n}=1.
\label{74c}
\end{equation}
The resonant states satisfy the Schr\"{o}dinger equation of the problem $[\kappa_n^2-H]u_n(r)=0$ with outgoing boundary conditions $u_n(0)=0$,\quad $[du_n(r)/dr]_{r=a}=i\kappa_nu_n(a)$,. The complex energy eigenvalues are $\kappa_n^2= E_n=\mathcal{E}_n-i\Gamma_n/2$, where $\mathcal{E}_n$ yields the resonance energy of the decaying fragment and $\Gamma_n$ stands for the resonance width,  which yields the lifetime (recalling that $\hbar=1$) $\tau_n=1/\Gamma_ n$ of a given resonance level. The lifetime of the system is defined by the longest lifetime, \textit{i.e.}, the shortest width. The complex poles $\kappa_n=a_n-ib_n$ are distributed along the third and fourth quadrants of the complex $k$ plane in a well known manner \cite{newton}. Notice that writing $G^+(r,r\,';k)=
G^+(r',a;k) \exp [ik(r-a]$ one may get a resonant expansion for $r'<a$ with $r >a$.
The above representation for $G^+(r,r\,';k)$ satisfies the closure relation for resonant states \cite{gc10},
\begin{equation}
{\color{red}
\frac{1}{2}}\sum_{n=-\infty}^{\infty} u_n(r)u_n(r\,')=\delta(r-r\,'),\quad  (r,r')^{\dagger} \leq a
\label{9y}
\end{equation}
and the sum rules,
\begin{equation}
\sum_{n=-\infty}^{\infty} \frac{u_n(r)u_n(r\,')}{\kappa_n}=0, \quad  (r,r')^{\dagger} \leq a,
\label{9yy}
\end{equation}
and
\begin{equation}
\sum_{n=-\infty}^{\infty} u_n(r)u_n(r\,')\kappa_n=0,  \quad  (r,r')^{\dagger} \leq a.
\label{9yyy}
\end{equation}

The above results permit to write the retarded Green's function as \cite{gc10}
\begin{equation}
g(r,r';t)=\sum_{n=-\infty}^{\infty}
\left \{ \begin{array}{cc}
u_n(r)u_n(r')M(y^\circ_n), & \quad  (r,r')^{\dagger} \leq a \\[.3cm]
u_n(r')u_n(a)M(y_n), & \quad r'< a, \,\, r> a,
\end{array}
\label{b6}
\right.
\end{equation}
where the functions $M(y_n)$, the so called Moshinsky functions, are defined as \cite{gc10}
\begin{equation}
M(y_n)=\frac{i}{2\pi}\int_{-\infty}^{\infty}\frac{{\rm e}^{ikr}{\rm e}^{-ik^2t}}{k-\kappa_n}dk=
\frac{1}{2}{\rm e}^{(imr^2/2 t)} w(iy_n),
\label{16c}
\end{equation}
$y_n={\rm e}^{-i\pi /4}(1/2t)^{1/2}[(r-a)-2 \kappa_nt]$, and the function $w(z)=\exp(-z^2)\rm{erfc(-iz)}$ stands for the Faddeyeva or complex error function \cite{abramowitz} for which there exist efficient computational tools \cite{poppe}. The argument $y_n^{\circ}$ of the functions $M(y_n^0)$ in (\ref{b6}) is that of $y_n$ above with $r=a$.

In what follows the discussion will be concerned with two quantities of interest in decaying problems: The survival probability $S(t)$ that yields the probability that at time $t$ the system remains in the initial state and the nonescape probability $P(t)$ that provides the probability that at time $t$ the particle still remains within the confining region of the potential. The survival probability follows immediately from the expression for the survival amplitude
\begin{equation}
A(t) = \int_0^a \Psi^*(r,0)\Psi(r,t) \,dr, \quad S(t)=|A(t)|^2,
\label{90}
\end{equation}
and the nonescape probability reads
\begin{equation}
P(t) = \int_0^a \Psi^*(r,t)\Psi(r,t) \,dr.
\label{91}
\end{equation}
One see that both quantities require only of the solution along $r \leq a$.
Hence inserting the first equation of (\ref{b6}) into equation (\ref{1s}) yields for the time-dependent decaying solution the exact expression
\begin{equation}
\Psi(r,t)=\sum_{n=-\infty}^{\infty}C_n(r)u_n(r)M(y_n^{\circ}),\qquad r< a,
\label{3s}
\end{equation}
where the coefficients $C_n$ are  defined as
\begin{equation}
C_n=\int_0^a \Psi(r,0) u_n(r) dr.
 \label{3c}
\end{equation}

Equation (\ref{3s}) may be used to study the short-time behavior of the survival probability. This will no be discussed here and refer the interested reader to ref. \cite{cgc12}, where it is shown that in general the
short-time behavior of the survival probability S(t ) has a dependence on the initial state and may behave as either $S(t)=1-\mathcal{O}(t^{3/2})$ or  $1-\mathcal{O}(t^{2})$. This might be of some interest because there is the widespread opinion that the short-time behavior of $S(t)$ must be quadratic.

Since the potential is real, it follows  from time-reversal invariance that $u_{-n}(r)=u_n^*(r)$ and $\kappa_{-n}=-\kappa_n^*$ \cite{rosenfeld61} that allows to express equation (\ref{3c}) as a sum running from $n=1$ to $\infty$.  Also, using some properties of the Faddeyeva function allows to write the Moshinsky function for the poles lying on the fourth quadrant as \cite{abramowitz,gc10},
\begin{equation}
M(y_{n}^{\circ})= {\rm e}^{-i \kappa_n^2 t} - M(-y_{n}^{\circ}).
\label{3xx}
\end{equation}
Substitution of equation (\ref{3xx}) into  equation (\ref{3s}) leads to the expression
\begin{equation}
\Psi(r,t)=\sum_{n=1}^{\infty}C_n(r)u_n(r){\rm e}^{-i \kappa_n^2 t} -\sum_{n=1}^{\infty} I_n(r,t),   \quad r< a,
\label{3s1}
\end{equation}
where $I_n(r,t)$ stands for the non-exponential contribution,
\begin{equation}
I_n(r,t)= [C_nu_n(r)M(-y_{n}^{\circ})-C_{-n}u_{-n}(r)M(y_{-n}^{\circ})],
\label{3xz}
\end{equation}
where the argument $y_{-n}^{\circ}$  is similar to that of $y_{n}^{\circ}$, defined above, with $\kappa_n$ substituted by $\kappa_{-n}=-\kappa_n^*$ and $C_{-n}$ follows from equation  (\ref{3c}) by recalling that $u_{-n}(r)=u_n^*(r)$. At long times these two functions exhibit an inverse power of time behavior.
In particular, for a real initial state, the decaying solution behaves along the exponential and long-time regimes as \cite{gc10},
\begin{equation}
\Psi(r,t) \approx \sum_{n=1}^{\infty} C_nu_n(r){\rm e}^{-i\mathcal{E}_nt}{\rm e}^{-\Gamma_nt} -
i \eta\,{\rm Im} \left\{\sum_{n=1}^{\infty}\frac{C_nu_n(r)}{\kappa_n^3} \right\} \,\frac{1}{t^{\,3/2}};\,\,r \leq a,
\label{3b}
\end{equation}
where $\eta=1/(4\pi i)^{1/2}$.

Assuming that the initial state $\Psi(r,0)$ is normalized to unity, there is an interesting relationship that follows from the closure relation (\ref{9y}),
\begin{equation}
{\rm Re}\sum_{n=1}^\infty \left\{ C_n \bar{C}_n\right\}= 1,
 \label{9z}
\end{equation}
where
\begin{equation}
{\bar C}_n=\int_0^a \Psi^*(r,0) u_n(r) dr.
\label{3e}
\end{equation}
Equation (\ref{9z}) indicates that although ${\rm Re}\,\{C_n{\bar C}_n\}$ cannot be interpreted as a probability, since in general is not a positive quantity, nevertheless it represents the `strength'  or `weight' of the initial state in the corresponding resonant state. If this has a value close to unity one may ignore the rest of the coefficients in the expansion for $\Psi(r,t)$. One may see the coefficients  ${\rm Re}\,\{C_n{\bar C}_n\}$ as some sort of quasi-probabilities \cite{halliwell}. This deserves further study.

It is worth mentioning that there is another route to analyze the long-time behavior of $g(r,r';t)$ which follows by closing the Bromwich contour mentioned above along a straight line $C_l$ that is $45^{\circ}$ off the real axis and goes through the origin as discussed in ref. \cite{gcmv07,gc10} that will not be considered here.

\section{Decay of two identical noninteracting particles}

In the case of a system of identical  non interacting particles, it is known that the Hamiltonian $H$ must be symmetric under the permutation of the indices of the particles so the exchange operator and $H$ necessarily commute. Thus, it is enough to impose the appropriate symmetry/antisymmetry on the initial state $\Psi(y_1,y_2,0)$ since symmetry is conserved as time evolves. Hence, the time evolution for decay of two identical particles may be written as
\begin{equation}
\Psi(\mathbf{r},t)= \int_0^a {\!\int_0^a {\!g(r_1,y_1,t)g(r_2,y_2,t)\Psi(\mathbf{y},0)\,\mathrm{d}y_1}\,\mathrm{d}y_2},
\label{eq:Psi_2_parts}
\end{equation}
where $\mathbf{r}$ and $\mathbf{y}$ denote, respectively,  $(r_1,r_2)$ and  $(y_1,y_2)$.

A simple choice, which corresponds to a symmetric state, is given by the product
of single particle states $\psi_\alpha(y_1)$ and $\psi_\alpha(y_2)$, with $\alpha$ denoting the state,
\begin{equation}
\Psi(\mathbf{y},0)=\psi_\alpha(y_1)\psi_\alpha(y_2).
\label{fac_sim}
\end{equation}
Substitution of (\ref{fac_sim})  into (\ref{eq:Psi_2_parts}), using the first equation in (\ref{b6}), yields the \textit{factorized symmetric} state
\begin{equation}
\Psi(\mathbf{r},t)=  \left(\sum_{p=-\infty}^\infty C_{p,\alpha}u_p(r_1)M(z_p)\right)
\left(\sum_{q=-\infty}^\infty C_{q,\alpha}u_q(r_2)M(z_q)\right)
\label{fac_Psi_exact}
\end{equation}
where $C_{n,\alpha}$, with $n=p,q$, is given by
\begin{equation}
C_{n,\alpha}=\int_0^a {\!u_n(y)\psi_\alpha(y)\,\mathrm{d}y}.
\label{coefs}
\end{equation}
Another choice for the initial  state consists of the linear combination of single-particle states $\psi_s(y_1)$
and $\psi_s(y_2)$.  Here, $s=\alpha,\beta$ refers to the possible states of the two particles, and hence we may write
\begin{equation}
\Psi(\mathbf{y},0)=\frac{1}{\sqrt{2}}(\psi_\alpha(y_1)\psi_\beta(y_2)\pm \psi_\beta(y_1)\psi_\alpha(y_2)),
\label{eq:e_inicial}
\end{equation}
where respectively,  the plus sign refers to \textit{entangled symmetric} and the minus sign to
\textit{entangled antisymmetric} states.
Then, substitution of (\ref{eq:e_inicial})  into (\ref{eq:Psi_2_parts}), using the first equation in (\ref{b6}), yields
\begin{equation}
\Psi(\mathbf{r},t)=  \frac{1}{\sqrt{2}}\sum_{p,q=-\infty}^\infty (C_{p,\alpha}C_{q,\beta} \pm C_{p,\beta}C_{q,\alpha})  u_p(r_1)u_q(r_2)M(y_p)M(y_q),
\label{twopsi}
\end{equation}
where the coefficients $C_{n,\beta}$  follow by replacing $\alpha$ for $\beta$ in (\ref{coefs}).
It is worth recalling that the coefficients $\{C_{n,s}\}$, which involve only single-particle states,
fulfill the relationship \cite{gc10}
\begin{equation}
{\rm Re} \left( \sum_{n=1}^\infty C_{n,s}{\bar C}_{n,s} \right )=1,
\label{sumrule}
\end{equation}
where ${\bar C}_{n,s}$ is defined as (\ref{coefs}) with $\psi_s(y)$ substituted by $\psi^*_s(y)$. Hence for real
initial states,  ${\bar C}_{n,s}=C_{n,s}$. Although the $C_{n,s}$ are complex and its real part may be negative,
they play a most relevant role in time-dependent expansions.

In a similar fashion as for the single-particle case, one may write the two-particle states
as a sum of exponential decaying contributions plus a long-time inverse power term. For the
\emph{entangled symmetric} state one may obtain from equation (\ref{twopsi})
\begin{equation}
\Psi(\mathbf{r},t)\approx\frac{1}{\sqrt{2}}\sum_{p,q=1}^\infty (C_{p,\alpha}C_{q,\beta}+C_{p,\beta}C_{q,\alpha})u_p(r_1)u_q(r_2){\rm e}^{-i(\mathcal{E}_p+\mathcal{E}_q)t}
{\rm e}^{-\frac{1}{2}(\Gamma_p+\Gamma_q)t}-A\frac{1}{t^3}.
\label{psis}
\end{equation}
and, similarly, for the \emph{entangled antisymmetric} state,
\begin{equation}
\Psi(\mathbf{r},t)\approx\frac{1}{\sqrt{2}}\sum_{p,q=1}^\infty (C_{p,\alpha}C_{q,\beta}-C_{p,\beta}C_{q,\alpha})u_p(r_1)u_q(r_2){\rm e}^{-i(\mathcal{E}_p+\mathcal{E}_q)t}
{\rm e}^{-\frac{1}{2}(\Gamma_p+\Gamma_q)t}+ B\frac{1}{t^5}.
\label{psia}
\end{equation}
where in the above expressions $A$ and $B$ are constants. These quantities have been obtained explicitly for the model of a $delta$-shell interaction \cite{gcm11}. The relevant point is the distinct long-time inverse power behavior of  \emph{symmetric} and \emph{antisymmetric} decaying states, which seems to possess a general character \cite{delcampo11}.

The survival amplitude of a two-particle system is defined as
\begin{equation}
\label{eq:def_amplit_A}
A(t)=\int_0^a {\! \int_0^a {\! \Psi^*(r_1,r_2,0)\Psi(r_1,r_2,t)\,\mathrm{d}r_1}\,\mathrm{d}r_2};
\end{equation}
hence, the survival probability is given simply by $S(t)=|A(t)|^2$.
The nonescape probability of a two-particle system is defined as
\begin{equation}
\label{eq:def_prob_P}
P(t)=\int_0^a {\! \int_0^a {\! |\Psi(\mathbf{r},t)|^2\,\mathrm{d}r_1}\,\mathrm{d}r_2}.
\end{equation}
Once $\Psi(\mathbf{r},t)$ is known, the calculation of $S(t)$ and $P(t)$ follows using the above expressions.

Lack of space prevents to illustrate  the analytical expressions derived above for calculations of the survival and nonescape probabilities for the single and two-particle cases. I refer the interested readers to go to refs. \cite{gcmv07} and \cite{gcm11}.

\section{Concluding remarks}

I hope I have been able to discuss the essential aspects of the formalism of resonant states for the time evolution of decay. It is surprising to see that over the years most of the attention has been for single particle decay. However, recent developments in the design and control of the number of atoms in ultracold traps and its interactions \cite{serwane11,pons12,rontani13} might lead to new and interesting findings involving the decay by tunneling of several particles.

\section*{Acknowledgments}
I acknowledge the partial financial support of UNAM-DGAPA-PAPIIT IN111814.
\section*{References}
\end{document}